\documentclass[nofootinbib, 
  reprint, amsmath,amssymb, aps, prl]{revtex4-1}
\usepackage{hyperref}
\def\Det{{\rm Det}}
 \usepackage{graphicx}% Include figure files
\usepackage{bm}% bold math
\def\O{\Omega}
\newcommand\re[1]{({\ref{#1}})} 
\def\be{\begin{eqnarray} }
\def\ee{\end{eqnarray}}  \def\IR{{\mathbb{R}}}
\def\IZ{{\mathbb{Z}}}  
 \def\no{\nonumber} \def\la{\label} 
\def\({\left(} \def\){\right)} \def\<{\left\langle\,} \def\>{\,
\right\rangle} \def\[{\left[} \def\]{\right]} \def\tr{{\rm tr} }
\def\hf{ {\textstyle{1\over 2}} } 
 \def\CS{{\cal S}}    
 \def\CI{{ \mathcal{ I} }}

       \def\CN{{ \cal N}}
      \def\CH{ {\cal H}}

     \def\p{\partial} \def\a{\alpha} \def\b{\beta} 
     \def\e{\epsilon}   \def\vp{\varphi}
     \def\th{\theta}  
     
  \def\d{\delta}

   \def\IO{{\mathbb{O}}}
     \def\ID{{\mathbb{D}}}
       \def\psu{{\mathfrak{psu} }}
  \def\Zh{ Zhukovsky\ }
  \def\ve{\varepsilon}
\def\C{\mathbf{C}}
 \def\CX{{ \mathcal{ X} }}
 \def\pf{\mathrm{pf}}
 \def\Pf{\mathrm{Pf}}
\def\K{\mathrm{K}}
\def\KKK{\mathbf{K}}
\def\JJ{ \bm{J}}
\def\II{ \bm{I}}
\def\CR{\mathcal{R}}
\def\CRR{ \bm{\mathcal{R}}}
\def\KK{ \bm{K}}

\begin{document}

\preprint{IPhT/t12/023}

\title{Determinant formula for the octagon form factor in $\CN=4$ SYM
}

\author{ Ivan Kostov$^a$, Valentina B. Petkova$^b$, Didina Serban$^a$}

\affiliation{%
	     $^a$Institut de Physique Th\'eorique, DRF-INP 3681,
	     C.E.A.-Saclay, F-91191 Gif-sur-Yvette, France }
 \affiliation{%
 $^b$Institute for Nuclear Research and Nuclear Energy,
 \\
Bulgarian Academy of Sciences, Sofia, Bulgaria}

\begin{abstract}

We compute to all loop orders correlation function of four heavy BPS
operators in $\CN= 4 $ SYM with special polarisations considered
recently by Frank Coronado.  Our main result is an expression for the
octagon form factor as determinant of a semi-infinite matrix.  We find
that at weak coupling the entries of this matrix are linear
combinations of ladder functions with simple rational coefficients and
give the full perturbative expansion of the octagon.
 \end{abstract}

\pacs{Valid PACS appear here}

\maketitle

\section{\label{sec:intro}Introduction}

The discovery of integrability in the planar $\CN=4$ SYM
\cite{Minahan:2002ve} initiated a `worldsheet' approach powered by the
analytic tools developed for two-dimensional solvable models.  In this
approach a single-trace operator is described as a state of a
two-dimensional field theory compactified on a circle.  The state
consists of a set of physical excitations on the top of a ground state
associated to a half-BPS operator.  By gauge-string duality, this is
also a closed string in the $AdS_5\times S^5$ background.

The full spectrum of such operators has been obtained for any value of
the gauge coupling applying the integrability techniques related to
the Thermodynamic Bethe Ansatz
\cite{Integrability-overview-2012,Gromov:2013pga, Gromov:2014caa}.
The computation of the OPE structure constants needed a new
theoretical input.  It came with the `hexagon proposal' of Basso,
Komatsu and Vieira \cite{BKV1}.  The authors of \cite{BKV1} proposed
to split the worldsheet of a three-point function into two hexagonal
patches, each containing a curvature defect.  The observables
associated with the two hexagons are special form factors which can be
computed using the symmetries of the theory.  The prescription using a
`hexagonalisation' of the worldsheet was then extended to the case of
the four-point functions \cite{Fleury:2016ykk,
Eden:2016xvg,Fleury:2017eph} and to non-planar corrections
\cite{Bargheer:2017nne, Bargheer:2018jvq}.  The hexagons are glued
back by inserting complete sets of virtual states in the intermediate
channels.

The contribution of virtual particles in the spectrum of `heavy'
operators ({\it i.e.} with large dimensions) is suppressed in the weak
coupling limit.  This is also the case for the three-point functions
of such operators.  In the strong coupling limit the virtual particles
cannot be neglected anymore, and in the cases amenable to analytical
treatment their contribution is expressed in terms of Fredholm
determinants \cite{JKKS2}.

\begin{figure}
         \centering
	 \begin{minipage}[t]{0.7\linewidth}
            \centering
            \includegraphics[width=5.2 cm]{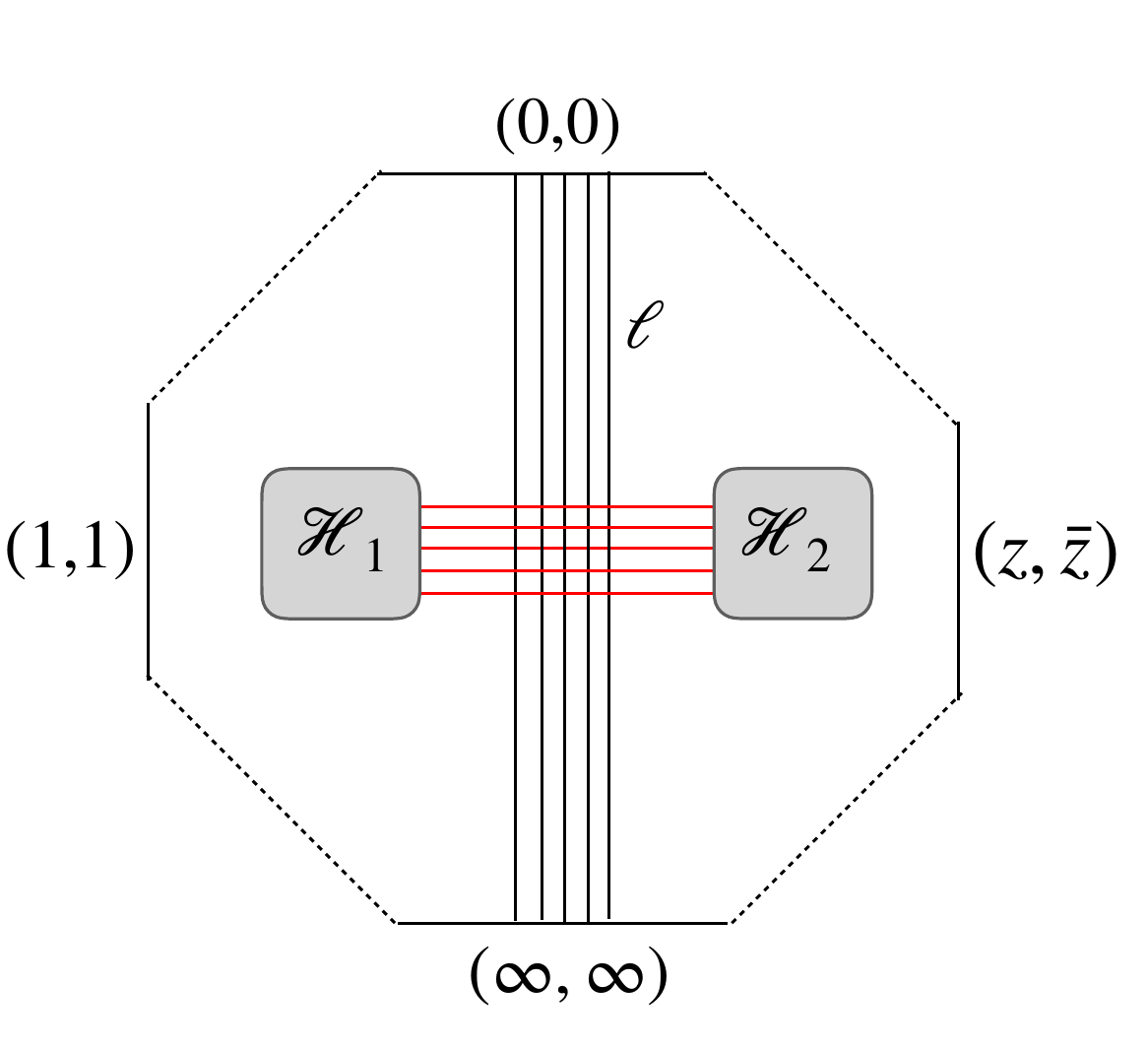}
 \caption{ \small A sketch of the octagon $\IO_\ell$.  The red lines
 symbolise the mirror particles propagating between the two hexagons,
 each one characterised by a rapidity $u$ and a bound state number
 $a$.  The two hexagons are separated by a  `bridge'  of length
 $\ell$.  }
  \label{fig:Octagon} 
         \end{minipage} 
            \end{figure}

In the computation of the four-point functions of heavy operators
by hexagonalisation, the virtual particles are not suppressed at weak
coupling anymore \cite{Fleury:2016ykk} and the evaluation of their
contribution represents a challenge.  Recently, Frank Coronado
obtained some remarkable results for the four-point functions of heavy
half-BPS operators with particular polarisations of the R-charges
\cite{Coronado:2018ypq,Coronado:2018cxj}.  In that configuration, the
four-point function factorises into sum of products of the so called
octagon form factors, or {\it octagons}.  An octagon is obtained by
gluing together two hexagons by inserting a complete set of virtual
particles.  The Boltzmann weights of the virtual particles depend on
the coordinates and the $R$-charge polarisations of the two hexagons,
as well as on the length $\ell$ of the `bridge' composed of tree-level
propagators (the vertical lines in Fig.  \ref{fig:Octagon}).

The octagon was expressed in \cite{Coronado:2018ypq} as an infinite
series of non-singular contour integrals which can be evaluated by
residues.  It is claimed that full perturbative expansion of the
octagon can be recast as a multilinear combination of conveniently
normalised ladder integrals $f_1, f_2, ...$ \cite{USSYUKINA1993363},
see equation \re{deffk} for their definition,

\noindent
  \be
 \begin{aligned}
\la{defInOa} \IO _{\ell}& =1+\sum_{n=1}^{\infty} \ \CX_n \
\sum_{J=n(n+\ell)} ^\infty \ g^{2J } \\
& \times  
\!\!\! 
\
\sum_{  j_1+...j_n=J} c_{j_1,\cdots j_n} 
\ f_{j_1} \cdots f_{j_n }
 \, ,\ \ \ 
 \end{aligned}
 \ee
where the dependence on the polarisations is carried by the factors
 \be
 \CX_n = \hf\( (\CX^+)^n + (\CX^-)^n\)
 \ee
and the coefficients $c_{j_1...j_n}$ are rational numbers to be
determined.  The conjectured form of the perturbative octagon, eq.
\re{defInOa}, is close in spirit to the result of Basso and Dixon
\cite{Basso:2017jwq} obtained for the fishnet limit of the $\CN=4 $
SYM \cite{Gurdogan:2015csr}\footnote{The integrability of the fishnet
Feynman graphs has been first established by A. Zamolodchikov
\cite{Zamolodchikov:1980mb}.}.
The analytic expression obtained in \cite{Basso:2017jwq} for the
fishnet has the form of a single determinant of ladders, while the
result of the bootstraping the Ansatz \re{defInOa} worked out in
\cite{Coronado:2018cxj} can be expanded in the minors of the
semi-infinite matrix
  \be \mathbf{f}_{\infty\times \infty}= [ f_{i+j+1}]_{i,j \ge \ell}\;
  \la{detf} \ee

In this Letter we report a formal solution for the octagon at finite
't Hooft coupling $g$ in the form of the Pfaffian of a semi-infinite
matrix, or equivalently as a determinant of the same matrix.
We confirm the Ansatz \re{defInOa} to high loop orders observing that
the resulting determinant is equivalent perturbatively to the
determinant of a simpler matrix whose elements are expressible in
terms of the ladder functions alone.  This leads to an analytic
expression for the coefficients in \re{defInOa}.
Moreover, the exact determinant representation for
finite $g$ opens the possibility to access analytically the four-point
function beyond the perturbative expansion.
 
We report here the results and the general logic of the derivation,
leaving the proofs to an extended paper.  We start in section II with
the representation of the contributions of the virtual particles as
Fredholm Pfaffians, outlined in \cite{Basso:2017khq}.  Then in section
III we  perform the sum over bound states and give the result of the
octagon as a single Fredholm Pfaffian, or the square root of a
Fredholm determinant.  In section IV we use a basis of Bessel
functions to transform the Fredholm determinant into the determinant
of a semi-infinite matrix.  In section V we derive the weak 
coupling expansion and show that it can be organised as a sum of
minors of a semi-infinite matrix of ladders.

\section{\label{sec:octagon} The octagon form factor}

In this section we remind the series expansion of the octagon as a sum
over virtual particles, which will be our starting point. 
% The
%dispersion relation for the physical particles $ E^2= {1\over 4} + 4
%g^2 \sin^2{p\over 2} $ is parametrised by the \Zh map $u\to x(u)$ with
% %
%\begin{align}\la{paramux}
%{u/ g} = x+{1/ x} .
%\end{align}
%%
The
virtual particles and their bound states propagate in the mirror 
dynamics and their energy and momentum are written, with the help of the shift operator $\ID=
e^{i \p_u/2}$, as
 \be\begin{split} \tilde p_a(u) &= \hf g \( \ID^a + \ID^{-a} \) (x-{1/
 x}), \\
  \tilde E_a(u) &= (\ID^a+\ID^{-a})\log x , \quad a=1,2,...,.
\end{split}
\ee
 where $x=x(u)$ is given by the \Zh map,
\begin{align}\la{paramux}
{u/ g} = x+{1/ x} .
\end{align}
%
%The change of sign in front of the one of the shift operators in the
%formula above is related to the so-called mirror transformation.
%The momentum and the energy of the physical particles or bound states
%of type $a$ are written, with the help of the shift operator $\ID=
%e^{i \p_u/2}$, as
% %
% \be
%\begin{split}
%p_a(u) &= - i (\ID^a - \ID^{-a})\log x, \\
% E_a(u)
% &=- {ig\over 2} \( \ID^a - \ID^{-a}\)
% (x- {1/   x})\, .
%\end{split}
%\ee
%%

Our goal is to evaluate the octagon with four physical and four mirror
edges with the corresponding BMN vacuum at each physical edge, as
shown schematically in Fig.  \ref{fig:Octagon}.  The octagon is
obtained by gluing the hexagons $\CH_1$ and $\CH_2$ along the common
edge $(0,0)$--$(\infty, \infty)$ by inserting a complete set of
virtual states $\psi$ with energies $E_\psi$.  Symbolically
 \be \IO_\ell = \sum_\psi \< \CH_2|\psi\> \, \tilde \mu_\psi \, e^{-
 E_\psi \ell} \, \< \psi|\CH_2\>.  \ee
where $\tilde \mu_\psi$ is a measure which will be detailed below.  A
state $\psi$ may contain any number of fundamental particles and their
bound states transforming in the anti-symmetric representations of
$\psu(2|2){\times }\psu(2|2) $.  An $n$-particle virtual state $\psi$
is completely characterised by the rapidities and the bound state
numbers $(u_j, a_j)$ of the individual particles $(j=1,..., n)$. 
% The
%virtual particles propagate in the mirror dynamics ($E\to i \tilde p,
%p\to i\tilde E$) whose energy and momentum are given by
%  %
% \be\begin{split} \tilde p_a(u) &= \hf g \( \ID^a + \ID^{-a} \) (x-{1/
% x}), \\
%  \tilde E_a(u) &= (\ID^a+\ID^{-a})\log x .
%\end{split}
%\ee
%%
%The change of sign in front of the one of the shift operators in the
%formula above is related to the so-called mirror transformation.

The four-point function depends on the cross ratios in the Minkowski
and in the flavour spaces, parametrised in \cite{Coronado:2018ypq} by
$z, \bar z, \a, \bar\a$.  Sometimes it will be more convenient to use
instead the phases $\xi, \phi, \vp, \th$ defined as
  \be 
  \la{defz} \begin{aligned} \la{exponentsdef} z &= {e^{-\xi+i\phi } } ,
  \quad\quad \bar z = {e^{-\xi - i\phi} } , \\
 \a& = { e^{\varphi -\xi  + i\theta}  } ,
  \ \ \
  \bar \a = { e^{\varphi -\xi - i\theta}} .
 \end{aligned}\ee
 Applying the hexagonalisation prescription, one obtains the series
 expansion for the octagon \cite{Coronado:2018ypq}
 \be
 \begin{aligned}
 \la{defOctserer} \IO_{\ell} &= {1\over 2} \sum_{\pm}
 \sum_{n=0}^\infty \sum_{a_1,..., a_n\ge 1} { (\CX^\pm/ \sqrt{z\bar
 z})^{n}\over n!} \int \prod_{j=1}^n {du_ j \over 2\pi   }\ \\
  &\times  {\sin (a_j\phi)\over \sin\phi}
 { \bm{\mu}}_{a_j}(u_j) 
    \prod_{j<k} \tilde H_{a_j, a_k} (u_j, u_k).
 \end{aligned}
 \ee
The different factors in the integrand are defined as follows.  The
symmetric bilocal factor ${\tilde H}_{ab}(u,v)$ is the product of
\be
\la{defK}
\begin{aligned}
K(u,v) &= {x(u)-x(v)\over x(u)x(v)-1}.  \end{aligned}\ee
with the four possible shifts of the arguments $u$ and $v$ in $\pm i
a/2$ and $\pm i b/2$ respectively, which we write symbolically as
\be
\begin{aligned}
\tilde H_{ab}(u,v)&= K(u,v)^{(\ID_u^a+\ID_u^{-a})(\ID_v^b+\ID_v^{-b})}
.
\end{aligned}
\ee
The local integration measure is
  \be\la{defmubf} \begin{aligned} { \bm{\mu}}_{a}(u)&= {1\over i g } \
  \ e^{-\tilde E_{a}(u) \ell } \ e^{ 2i \xi \, \tilde p_{a}(u) } \\
    \times &
 {1\over   \( x-x^{-1}\)^{ \ID^{a} \!+\ID^{-a} }}\ 
     { x (u+i{a\over
    2}) - \ x (u-i{a\over 2}) \over { x (u+i{a\over 2}) x (u-i{a\over
    2})-1} } .  \end{aligned}\ee
Finally the dependence on the polarisations is contained in the
factors  
  \be \CX^\pm = 2 (\cos \phi - \cosh( \varphi \mp i\th)) \ \sqrt{z\bar
  z} .  \ee

\section{The octagon as a Fredholm Pfaffian }

The expansion for the octagon, eq.  \re{defOctserer}, resembles the
grand partition function of a Coulomb gas of dipoles.  As it has been
first pointed out in \cite{Basso:2017khq}, the product of the bi-local
weights in the $n$-particle sector can be written as a Pfaffian of a
$2n \times 2n$ matrix, and the whole expansion as a sum of two
Fredholm Pfaffians \cite{2000math......6097R},
 \be
 \begin{aligned}
 \la{defInPf} \IO_{\ell} &=\hf \sum_{\pm} \sum_{n=0}^\infty
 {(\CX^\pm)^n\over n!} \sum_{a_1,...,a_n\ge 1} \int \prod_{j=1}^n \
 d\mu(u_j,a_j) \\
  \times & \pf \[ K ( u_j +i{\ve_j a_j/ 2}, u_k +i{\ve_k a_k/ 2}) \]
  _{^{\ve_j ,\ve_k=\pm 1}_{ j, k=1,\dots, n}}\qquad
  \\
   &= \hf\sum_{\pm} { \Pf\[ \JJ+ \CX^\pm \KK\]} .  \end{aligned} \ee
 In the last line $\JJ = [J^{\ve\d}]_{\ve, \d = \pm}$ is a $2\times 2$
 anti-symmetric matrix with non-zero elements $J^{+-}= - J^{-+}=1$,
 and $\KK$ is a $2\times 2$ anti-symmetric matrix kernel $ \KK=
 [K^{\ve\d}]$.  The kernel elements $K^{\ve\d}$ act in $\IR\times
 \IZ_+$ so that the integral in the rapidity $u\in\IR$ is accompanied
 by a sum over the bound state label $a\in \IZ_+$.  With the help of
 the shift operator $\ID$ the kernel $\KK$ can be written compactly as
\be \la{pfafker} \KK (u,a; v,b)= \[ K(u,v)^{\ID_u^{\ve a} \ID_v^{\d
b}}\]_{\ve, \d = \pm}.  \ee
The last factor in the integration measure \re{defmubf} is absorbed
into the Pfaffian and the rest gives the integration measure in
\re{defInPf}, which we write in the form
\be \la{measuredma}
\begin{aligned}
d\mu(u,a)&=  {1\over  i g \sqrt{z\bar z}}\ {du\over 2\pi  } \
{\sin(a\phi)\over \sin\phi}\, \O_\ell (u )^{\ID^{a} +\ID^{-a}} ,
  \end{aligned}
   \ee
with 
  \be \la{defOl}
   \begin{aligned}
    \O_\ell (u ) &\equiv {e^{ig \xi\, [x(u)-1/x(u)]}\over x(u)-1/x(u)}
    \ x(u)^{-\ell} .  \end{aligned} \ee
To compute the Fredholm Pfaffian we first express it as a square root
of a Fredholm determinant,
 \be
 \begin{split}\la{PfDet} 
 \Pf &[\JJ+\CX^\pm \KK]= \sqrt{\Det\[ \II -\CX^\pm \JJ \KK\]}=e^{ \CS
 ^\pm }
 ,
 \end{split}
 \ee
where $\II$ is the $2\times 2$ identity matrix.  Then we expand the
exponent $\CS^\pm$ as an infinite series of cyclic integrals/sums
\be \CS^\pm =- \sum_{n=1}^\infty {(\CX^\pm )^n\over 2n} \ \CI_n, \ee
where the $n$-th integral/sum reads (with $u_{n+1}\equiv u_1$ etc.)
   %
  %\ 
  \la{cyclicIn} 
   \begin{align}
\CI_n & = {1\over (ig\sqrt{z\bar z})^n} \sum_{a_1, ..., a_n\ge 1}
\sum_{\ve_1, ...  , \ve_n= \pm} \ \prod_{j=1}^n \int\limits
_{-\infty}^\infty \prod_{j=1}^n {du_j\over 2\pi i}\ \no \\
    &\times \prod_{j=1}^n {\sin(a_j\phi)\over \sin\phi} \O_\ell (u_j
    -i \ve_j {a_j\over 2})\ \O_\ell (u_j+ i\ve_j {a_j\over 2}) \no \\
  &\times \ve_j \, K(u_j - \ve_j {a_j\over 2} , u_{j+1} +i \ve_{j+1}
  {a_{j+1}\over 2} ) .
    \end{align}
 The sum over the bound state labels can be taken into account by the
 difference operator $[\cos\phi - \cos \p_u ]^{-1}$ applied to the
 simpler kernel \re{defK}.  We will give the details of the
 computatioon in a forthcoming work \cite{Longpaperoctagon}.

\section{Discrete basis}

To render the Pfaffian representation useful we have to find a way to
perform also the multiple integrals over the rapidities in the
expansion of $\CS^\pm$.  The formula we obtain is an
infinite-dimensional version of the Pfaffian integration theorem
\cite{Akemann:2007wa,1751-8121-40-36-F01}.  We expand the scalar
kernel \re{defK} with $ |x|>1$ and $|y|>1 $ as a double series
\be\la{basisK} K(u,v)= \frac{x-y}{x y-1}= \sum _{m, n=0}^{\infty }
x^{-n} \ \mathrm{C}_{nm} \ y^{-m} , \ee
where  
\be \la{defC} \mathrm{C}_{nm}=\d_{n+1, m }- \d_{n, m+1}, \qquad m,n\ge
0.  \ee
Substituting  \re{basisK} in the $n$-fold cyclic integral,
we achieve that   the latter  decouples into a sum of products of simple
integrals 
  \be
 \la{defKker}
\begin{aligned}
 \K _{mn} &= \sum_{\ve=\pm}{1\over 2i g\sqrt{z\bar z}} \int {du\over
 2\pi } \ \O _{\ell+n}(u - \ve i \e) \\
 &\times { \ve \over  \cos \p_u -\cos\phi   } \ \O _{\ell+m}(u+ \ve i
 \e) 
   \end{aligned}\ee
 with  $m,n\ge 0$.
The arguments on the rhs are displaced from the real axis by small
amounts $\pm i \e$ with $\e>0$ to avoid the \Zh cut.  Introducing the
semi-infinite matrices $\C$ and $\KKK$ with matrix elements given
respectively by \re{defC} and \re{defKker}, the exponents $\CS ^\pm$
take the form
  \be \la{defCAbb} \begin{aligned} &\CS ^\pm= - \hf \sum_{n=1}^\infty
  {( \CX^\pm)^{n}\over n} \tr (\C \KKK)^n.
    \end{aligned}
    \ee

 The matrix elements of $\KKK$ can be evaluated by passing to Fourier
 space, after which the integral in $u$ can be taken and results in a
 product of two Bessel functions.  The remaining integral in the
 Fourier variable $t$ is
\be\la{integralforK}
\begin{aligned}
  \K _{mn} &= {g\over 2 i\sqrt{z\bar z}} \int _{ |\xi|}^\infty d t {
  \( i\sqrt{ t+\xi \over t-\xi}\)^{ m-n }\!\!\!\!  - \( i\sqrt{ t+\xi
  \over t-\xi}\)^{ n-m } \over  \cos\phi-\cosh t  }\qquad \\
 &\times J_{m+\ell}(2g\sqrt{ t^2-\xi^2})
 J_{n+\ell}(2g\sqrt{ t^2-\xi^2}).  
  \quad
  \end{aligned}
 \ee 
The transformation to a discrete basis allowed us to write the
Fredholm Pfaffian \re{defInPf} as a square root of the determinant of
a semi-infinite matrix
  \be\begin{aligned} \la{PfOctbispfa} \IO_{\ell}(g, z, \bar z, \a,
  \bar\a)&= \hf\sum_{\pm} \sqrt{ \det\[ 1- \CX^\pm \C \KKK\]} .
              \end{aligned}
    \ee
Eqs.  \re{integralforK} and \re{PfOctbispfa} give a formal solution
for the octagon for any value of the gauge coupling $g$.

\section{Determinant formula for the perturbative octagon }

In the rest of this letter we will focus on the weak coupling
expansion of the octagon.  We will demonstrate the efficiency of our
formulas \re{defCAbb} or \re{PfOctbispfa}, each of which can be used
to reproduce and extend to virtually any loop order the results of
\cite{Coronado:2018ypq,Coronado:2018cxj}.

 The perturbative expansion of the matrix elements of $\KKK$, eq.
 \re{integralforK}, can be expressed in terms of the (conveniently
 normalised) ladder Feynman integrals \cite{USSYUKINA1993363}.  For
 the $1\times k$ ladder integrals we will use the notations and the
 normalisation of \cite{Coronado:2018cxj}\footnote{Up to a factor $v=(1-z)(1-\bar z)$: $f_n^{\text{Coronado}} = - v\,  f_n^{\text{here}}$. }   
\be \la{deffk}
\begin{aligned}
f_k &=   \sum_{j=k}^{2k} {(k-1)!  \ j!  \over (j-k)!  (2k-j)!} \\
 &\times | 2\xi|^{2k-j} \ {\mathrm{Li}_{j}(z) - \mathrm{Li}_{j}(\bar
 z)\over z-\bar z} ,
\end{aligned}
\ee
with $ 2 \xi = - \log z\bar z$ defined in   \re{defz}.  More precisely, we found that  %the
$\K_{ij}$ as functions of $g, z,\bar z$ are spanned by $\{f_m \xi^n
g^{2m +n}\}_{m\ge \ell+1\,,n\ge 0}$.

Substituting this expansion in moments \re{defCAbb} one can easily
reconstruct, with the help of Mathematica, the perturbative series for
the octagon.  Remarkably, all positive powers of $\xi$ cancel and the
result comes out in the form \re{defInOa}.  This is for now an
empirical observation which awaits its analytic proof.  It means that
the superfluous   %, besides that in the ladders $f_j$} 
$\xi$-dependence can be eliminated by a unitary transformation.
We can thus simplify drastically the computation by replacing the
matrix $\KKK $ in the traces \re{defCAbb} with the matrix $\KKK^\circ$
obtained by truncating the expansion of $\KKK $ to the subset $\{f_m
g^{2m}\}_{m\ge \ell+1}$.

The matrix elements of $ \KKK^\circ$ whose indices have the same
parity vanish.  This property, satisfied also by the constant matrix
$\C$, implies that $ \det[1-\CX^\pm\, \C\KKK^\circ]$ is equal to the
square of another determinant, $\det[1 + \CX^\pm\, \CRR]$, with the
matrix elements of $\CRR$ given by
\be \la{defR} \CR_{jk} = - \K^\circ_{2j+1, 2k}+ \K^\circ_{2j-1, 2k} .
\ee
Now the square root in \re{PfOctbispfa} gets resolved and the
determinant representation of the octagon simplifies to
   \be\begin{aligned} \la{PfOctbisdet} \IO_{\ell} &= \hf\sum_{\pm} {
   \det\[ 1+ \CX^\pm \, \CRR\]} .
 \end{aligned}
    \ee
 The perturbative series for $\CRR$ is
\be
\begin{aligned}
\la{seriesR} &\CR_{ij} = \sum_{ p= \max ( i+j+\ell, {1+ j+\ell})}^\infty
(-1)^{p-\ell} (2 p-1)!  \\ \times & \frac{ 2p(2 i+\ell) \ - \ (p - j )
(p+j+\ell)\ \delta _{i,0} }{ \prod\limits_{\ve = \pm} ( p + \ve (i-
j))!  ( p + \ve (i+j+\ell))!  }\ f_p \ g^{2p}.  \end{aligned}\ee

Eqs.  \re{PfOctbisdet}-\re{seriesR} give the all-loop perturbative
solution for the octagon.  For actual computations it is convenient to
truncate the semi-infinite matrix $\CRR$ to an $N\times N$ matrix
 \be \CRR_{N\times N}= \[\CR_{ij}\]_{0\le i,j\le N-1} \ee
and use the approximation formula
\be\la{octagonfinalpR} \IO_{\ell =0}=\hf \sum_{\pm} \det(1+ \CX^\pm
\CRR)_{_{N\times N}} + o(g^{^{ 2N(2N+\ell)}}).  \ee
For example, with $ N=3$ the determinant in \re{octagonfinalpR} 
gives the result of Coronado for $\ell=0$ up to $g^{12}$ terms:
\be\begin{aligned}\no &\IO_{\ell=0} =1 \\
 &+ \CX_1 \ \( f_1 g^2-f_2 g^4 + \hf f_3 g^6 -\textstyle{5\over 36}
   f_4 g^8 %+   \textstyle{7\over 288}  f_5  g^{10} 
%   -\frac{7}{2400} f_6 g^{12}
%  +\frac{11 }{43200} f_7 g^{14} 
  + \frac{7 f_5 g^{10}}{288}- ...
\) 
\\
 &+ \CX_2 \ \( \textstyle{f_1 f_3-f_2^2\over 12} g^8- \textstyle{f_1
 f_4 -f_2 f_3\over 24} g^{10}+
%\textstyle{1\over 720} 
%(7 f_1 f_5-9 f_3^2+2 f_2 f_4)g^{12}+... 
... \) 
\\
 &+ \CX_3 \( \textstyle{\left(f_1 f_5 f_3-f_3^3+2 f_2 f_4 f_3-f_1
 f_4^2-f_2^2 f_5\right)\over 34560} g^{18} + ...\) \\
&+\dots
\end{aligned}
\ee

In \cite{Coronado:2018cxj}, the octagon was expanded in a basis of
minors of the matrix \re{detf}.  In particular, the lowest loop order
$n$-particle contribution is proportional to the determinant of the
matrix \re{detf} restricted to the first $n$ rows and columns,
 \be
 \begin{split}
& \IO_{\ell} = \sum_{n=0}^\infty \CX_n \ g^{2n(n+\ell)}\ \(
I_{n+\ell,n} + o(g^2)\) \, , \\
& I_{n+\ell,n}(z,\bar z) = {{\det\(\[f_{i+j+\ell+1} \]_{i,j=0,...,
n-1}\)\over \prod_{i=0}^{n-1} (2i+\ell)!(2i+\ell+1)!}} \, .
\end{split}
 \ee
The lowest term $I_{n+\ell,n}$ is exactly the expression obtained by
Basso and Dixon \cite{Basso:2017jwq} for the Feynman integral for an
$(\ell+n)\times n$ fishnet.  One can recognise this pattern in Fig.
\ref{fig:Octagon} where $n$ virtual particles cross $\ell$ physical
particles.  Possibly an interpretation of the higher loop terms in terms of planar Feynman graphs also exists.
  
This expansion in the minors of the matrix of ladders $
\mathbf{f}_{\infty\times \infty}$ is compatible with our determinant
representation \re{octagonfinalpR}, which can be written as a sum over
all minors of the matrix $\CRR$, eq.  \re{seriesR},
 \be \la{minoresR} \IO_{\ell}= \sum_{n=0}^\infty \CX_n \sum_{^{0\le
 i_1< ...<i_n} _{0\le j_1 <...<j_n}} \det \(\[ \CR_{i_\a
 j_\b}\]_{\a,\b=1,..., n}\).\ \ \ \ee
 Since the matrix elements of $\CRR$ behave as $\CR_{ij} \sim
 g^{2(i+j+\ell)}+$higher powers of $g$, the lowest loop order
 contribution is given by the term $\sim g^{2 n(n+\ell)}$ of the
 $n\times n$ minor with $i_\a= \a+\ell -1, j_\b = \b+\ell -1$, which
 is exactly $I_{n+\ell, n}$.

It is straightforward to extract from \re{minoresR} the analytic
formula for the coefficients in the expansion in the Steinmann basis
of minors, but this would go beyond the scope of this short note.
     
 The representations \re{defCAbb} or \re{PfOctbispfa} could give for
 the first time analytic access to the correlation functions at finite
 $g$.  Remarkably, all the dependence on the gauge coupling is
 contained in a single integral \re{integralforK}.  We would like to
 address the subtle problem of the computation of the octagon at
 finite $g$ in a future work.

\begin{acknowledgments}
  
\noindent The authors are obliged to Till Bargheer, Frank Coronado and
Pedro Vieira for discussions and useful exchanges.  This research is
partially supported by the Bulgarian NSF grant DN 18/1 and by the
bilateral grant STC/Bulgaria-France 01/6, PHC RILA 2018 N$^\circ$
38658NG.
\end{acknowledgments}

% % \footnotesize
%% 
% \bibliography{/Users/vani/Dropbox/Valya/ABib-dropbox.bib}
%% \bibliography{ABib}
%  \bibliographystyle{/Users/vani/Files/PAPERS/PAPERSLIBRARY/utcaps}
% \end{document}
 
\providecommand{\href}[2]{#2}\begingroup\raggedright\endgroup

\end{document}